# Detection and Prediction of Infectious Diseases Using IoT Sensors: A Review

Mohammad Meraj[1], Surendra Pal Singh[1], Prashant Johri[2], Mohammad Tabrez Quasim[3]
*1. CS & IT Department ,NIMS University, Rajasthan Jaipur, India*
*2.School of Computer Sciences & Eng. GalgotiaUniversity,GrNoida,India*
*3.College of Computer &IT,University of Bisha ,Saudi Arabia*

ABSTRACT: An infectious kind of disease affects a huge number of human beings. A lot of investigation being conducted throughout the world. There are many interactive hardware platform packages like IoT in healthcare including smart tracking, smart sensors, and clinical device integration available in the market. Emerging technology like IoT has a notable ability to hold patients secure and healthful and also enhance how physicians supply care. Healthcare IoT also can bolster affected person pride by permitting patients to spend more time interacting with their medical doctors due to the fact docs aren't as taken with the mundane and rote aspects of their career. The most considerable advantage to IoT in healthcare is that it supports doctors in undertaking extra significant clinical work in a profession that already is experiencing a worldwide professional hard work shortage. This paper investigates the basis exploration of the applicability of IoT in the healthcare System.

## 1. INTRODUCTION

### 1.1 *IoT Impact on Healthcare*

Interactive hardware platform packages like IoT includes customized sensors that are designed to sense and respond to typical behaviors or situations, offering high performance( Khan,2020a; Quasim, 2019). As in healthcare, a deep investigation has been performed on the various Infectious diseases. The complex data gathering with the secure connection has enabled the use of blood glucose meters, fertility sensors and flu monitors, either by individual consumers or at medical institutions; ice sensors ensure that temperatures are insufficient to protect Notify technicians when vaccines and injections are available; Remote monitoring equipment enables patients to recover at home, while still benefiting from health care professionals who monitor the situation(Jaafar,2020). In terms of infection prevention and control, the power of the Internet of Things can be categorized into three different zones:
- Testing Parameters
- Performance behaviors
- Interaction with data

### 1.2 *Analysis of Human Behaviors*

The spread of diseases in health care is a research zone for researchers. Unfortunately, there has been a lack of evidence-based methods.

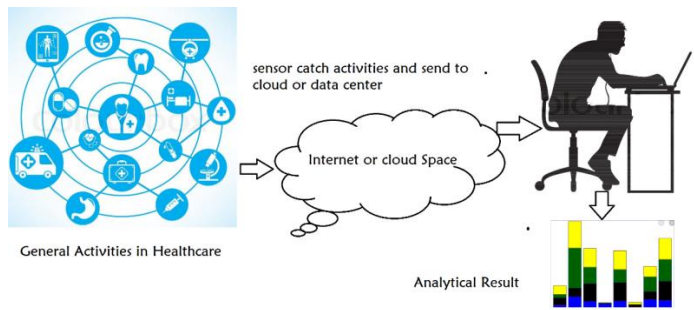

Figure 1. Smart Disease Surveillance Based on IoT

Analysis of Human behavior is very essentials for the right kind of investigation through sensors and the data gathered from the sensors.

1.3 *Collaboration with IoT*

Connect to machines and anything else that's never happened in the Internet of Things. The opportunity to get involved and collaborate locally and globally through targeted, context-relevant stages and groups is exciting and intriguing. A Systemic Investigation infection control techniques introduced "Surveillance Syndrome" (Furness, 2016), which includes organized gathering and analysis based on the status of symptoms seen in hospitals and other medical centers. Signal recruitment is a faster way to detect a public health outbreak or infection in a specific institution than current methods pending lab results on the shelf. Internet-based reporting tools, easy to use in emergency rooms, nurse's centers and doctors' offices will allow the rapid collection of relevant information. Information and analysis will be more accessible, open up the interplay between machines and humans, and reduce response times and intervention strategies.

1.4 *IoT Monitoring and Disease*

The Internet of Things (Khan ,2020b)is changing lives is Johns Hopkins University's Global Clinical Health Education Center, which uses network reacts to patient participation and help doctors and nurses share real-time health information in the care setting. Johns Hopkins University used this model on the investigation of tuberculosis on pregnant women. With increasing skepticism of this outbreak, Healthcare devices could be the IoT device network to obtain highly sensitive data to identify the source of this devastating disease. If an outbreak is confirmed, the same network can be used or developed to provide the required medication, medical equipment, and other diagnostic tools.

1.5 *Real- time application of IoT*

It has various applications in real-time like smart homes,industrial monitoring, and Noncivilian Purposed monitoring, remote location monitoring, E-healthcare, etc(Sivaram,2020). The use of IoT devices has very cost-effective and reliable use in engineering applications also. The Device could support the existing operating system and can handle it through a monitor or any suitable display.

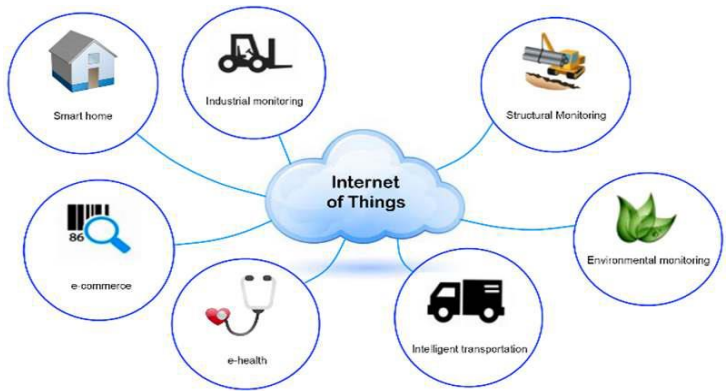

Figure 2. Application of IoT and Surroundings

with the rapid population growth of the elderly worldwide, the health care system is under tremendous pressure. To care patients it needs a variety of things like Hospital, Doctors, as well as more important checkup devices. This requires an explanation that reduces the problem on health care (McCue, 2015) system while continuing to provide patient health care. With the arrival of the smart devices and its adoption of jobs, it is extensively investigated as a possible explanation to shrink the burden on health care types. The aiding rehabilitation through continuous observations of progress and emergency health care (Furness, 2016).

2. RESEARCH BACKGROUND

(Matthew ,2015) Focusing on the general collection, analysis, and interpretation of local disease pattern data, which helps to identify outbreaks of major health-related symptoms. In developing countries like India, the performance of single surveillance systems is often hampered by data quality and availability. The Internet of Things (IoT) concept enables access to information about tagged objects or people by browsing the Internet address or data entry associated with a specific functional RFID with a visual function. In a "Smart recruitment" based on the Internet of Things, the smart device will process and send the required information to the spine network, which is located on a large server that keeps records of patients in each hospital. This spinal network will process and provide information to the Department of Health for faster, smarter, and easier to understand trends. As all information is provided through smart devices, this will help government agencies take the necessary steps without causing excessive delays.

(Verma ,2018)Demonstrate the theoretical framework of a mobile medical system that produces user diagnostic results (UDR) based on the life-balance results delivered by medical agents and other sensors. In addition, the formal model includes keywords, concepts, disease diagnostic methods, and generation awareness strategies. The main motivation for generating results with various diagnostic applications is to use different IoT measurement sets to analyze life better over a given period of time. In addition, the proposed SSIS is a patient method that can extract results from the data gathered by sensors for further health, observing, and delivered the specific task.

(Parthasarathy,2018 )proposed universal design created on the concept of IoT to process sensory evidence and is separated from the most important clinical limitations for joint pain. The most notable scientific parameter that confirms bone formation is related to ROC research assistance. Experimental results show that in the symptoms of arthritis and certain cardiovascular diseases, the levels of uric acid and C-reactive protein in the blood are averaged 8.50 and 3.50 mg / l, respectively. Human vehicle validation is used to isolate the onset of osteoporosis. In this article, with the help of TW calculations and UA and CRP levels in the blood, IoT devices are used to show the human development in a consistent way to use this IoT gadget. As many health care information has been identified as a result of the diagnosis of rheumatoid arthritis, the recommended work is very important.

(Laplante ,2017 )In this article, they propose a systematic way of defining smart gadgets for the E-smart health care system. They demonstrate and present related quality of work from the E-smart health care system especially the essential to regard care as a necessity. In this article, they present a formal framework for defining and later providing assistance in defining, developing, and using IoT in healthcare. This approach involves defining a common category of program types, classifying health care settings, and using a systematic way to describe the components of a particular use case. Using this approach to describe (e.g. explain) healthcare IoT can lead to stand-outs, reuse, collaboration, best practices, and more.

(Lopez-Barbosa,2016 ).Rapid discovery is an area of intense research that has attracted the efforts and interest of many fields and companies. However, the public interest in such devices has also grown, leading to major changes in the design and ideas of these methods of development and data generation. Therefore, future clinical trials should include communication with pre-existing technologies such as smartphones and networks, using automation of manufacturing. On the basis of this project, three key models have been analyzed based on various sensing technologies to highlight the development methods and tools for developing future clinical diagnostic equipment and data capture and application.

(Sattar ,2019) presented a paper for wound monitoring through IoT based system. They proposed a System develop on IoT based intelligent wound assessment for wound status. They used the ID3 algorithm as a decision tree for categorizing the attributes through MATLAB. This proposed system based on entropy and information gain for the splitting the tree and feature reduction.

(Liu,2017)This article introduces the origins and concepts on the impacts of smart devices and applications. This article outlines important issues related to IoT technology and service development. The most appropriate parameters for the application are suggested. Although they present the best applications for the Internet of Things, this article may not cover the challenges Internet of Things may face in future practical applications, and future research may begin in this regard.

(Muthu Kumar, 2019)Their work looks at this seemingly less moisture, but plays an important role in a person's well-being. Our work considered the beneficial effects of moisture storage and developed an IoT sensor-based module that monitors room temperature and updates the condition to room occupants. The system also regulates air-conditioners and air-conditioning in the living room to maintain optimal humidity levels between 45% and 55%. An upgraded module is an effective and very effective solution for hospitals because in such cases where the risk of an outbreak is high.

(Husain, 2015)In this article, they propose a human-based conceptual framework for health care for the elderly and disabled. The platform is designed to monitor the health of the elderly and disabled and give them an emergency response when their health is unfamiliar. They focus on three aspects:

(a) Content fraud (per person) from mobile devices;

(b) Instant Response

(c) Contextual mobile services to remote users.

(Zhu et.al 2020) illustrated how the infectious disease spread and turned up as a pandemic. They demonstrate that the IoT system becomes an important tool for healthcare centers to handle in a pandemic scenario. They found the IoT system is fast, user-friendly, and affordable to control the infectious disease.

(Ravi et.al 2019) explored the infectious disease like Malaria among humans. This is a quite long and difficult task for detection. As this problem is not much a life-threatening but may be converted in the serious problem. As the author found that few parameters like Blood sugar, body temperature, and heartbeat has significantly correlated with malaria. They proposed a model based on logistic regression to achieve maximum accuracy and speed in detection the process.

## 3. IMPLEMENTING EFFECTIVE INFECTION PREVENTION MECHANISMS

Based on this analysis, they can recommend the use of data from the IoT network to take preventive measures, or even decide whether the proposed control measures are being used effectively. For example, with the outbreak of influenza in the United States, IoT-enabled devices can track hand washing and virus usage, and collected data can be used to track its impact on influenza transmission in real-time. However, getting the job done right needs appropriate preparation and vigilant use of the right knowledge stages and implements. With the Internet of Things technology, it has become more realistic to actively identify and prevent the blowout of transferable infections.

## 4. EXPLORATION OF OUTCOME

In an assessment of this incipient technology is adding value to the sanitation systems of medical institutions, companies are advised to provide unlimited evidence of the effectiveness of their products. Merely discussing facts and statistics of health care-related illnesses (HAIs) or the ability of a product of smart devices that kill viruses is not enough. Prior database (Quasim,2013a; Quasim, 2013b) provide the right kind of help in the establishment of gadgets through IoT. Unless companies can provide evidence that their products have reduced overall HAIs over a period of time and space, cash-strapped health plans to slow down the use of IoT based smart application and devices. Technology can coordinate efforts to prevent infection by many hospitals through the spread of disease, including cleaning and disinfecting (people and machines) and hygiene practices by medical to their own treatments and help to reduce disease spreading and provides important messages to health care patients with a better decision-making tool.

## 5. CONCLUSION

The IoT devices are much capable to handle various things. The work with accuracy provides the IoT has a very power full device of present era. The multipurpose use of the device uplifts the various existing gadgets much capable to handle the complex task also. This paper is the exploration of various filed where the IoTs are working. This investigation is deeply focused on the use of IoT with the disease, especially in Infectious. This research found that the spreading of this disease may turn up as the epidemic. So the control of the Infectious disease is much-waiting kind of research investigation.